\documentclass[11pt,twoside,a4paper]{article}
\usepackage[a4paper, total={6in, 9in}]{geometry}

\usepackage[T1]{fontenc}
\usepackage[utf8]{inputenc}

\usepackage{amsmath, amsfonts, amsthm}
\usepackage{graphicx}
\usepackage{hyperref}
\usepackage{braket}
\usepackage{enumerate}
\usepackage{multirow}
\usepackage{adjustbox}
\usepackage{tikz}
\usetikzlibrary{quantikz}
\usepackage{soul}
\usepackage{color}
\usepackage{algorithm}
\usepackage{algpseudocode}
\usepackage{float}
\usepackage{caption}
\usepackage{subcaption}

\newtheorem{definition}{Definition}

\begin{document}

\begin{center}
{\Large
\textbf\newline\newline\newline{\textbf{\emph{NoTaDS}}: Distributed Scheduling of Quantum Circuits\\ with Noise and Time Optimization}
}
\end{center}

\begin{center}
Debasmita Bhoumik \textsuperscript{1,*},
Ritajit Majumdar \textsuperscript{2},\\
Amit Saha \textsuperscript{3},
Susmita Sur-Kolay\textsuperscript{1}
\\
\bigskip
\textsuperscript{1} Advanced Computing \& Microelectronics Unit,\\ Indian Statistical Institute, India\\
\textsuperscript{2} \emph{IBM Quantum}, IBM India Research Lab\\
\textsuperscript{3} Inria, Paris, and École Normale Supérieure (ENS), PSL University, Paris\\
* debasmita.ria21@gmail.com

\end{center}


\begin{abstract}
Quantum computers are currently noisy, particularly without error correction and fault tolerance. Methods like error suppression and mitigation are widely used to improve performance. Circuit cutting, which partitions a circuit into smaller subcircuits, can also reduce noise. In this paper, we propose an Integer Linear Program (ILP) based scheduler for optimizing subcircuit schedules on available hardware. The goal is to maximize overall fidelity and ensure each hardware does not exceed its predefined execution time. For 10-qubit circuits, our method achieves an average fidelity improvement of ~12.3\% and ~21\% with and without measurement error mitigation, respectively, even with minimal execution time. Additionally, we introduce a polynomial-time graph-theoretic scheduling method that matches the ILP scheduler's results when the number of subcircuits does not exceed the number of hardware units, each with minimal execution time. This noise and time-optimized scheduler represents a crucial step towards optimal quantum computing performance, especially with limited hardware access.
\end{abstract}

\section{Introduction}
\label{sec:introduction}

%
Quantum computers have been shown to be able to perform certain computations faster and/or more accurately than their classical counterparts \cite{shor1999polynomial, harrow2017quantum}. However, these algorithms assume the existence of fault-tolerant quantum computers, which are still not available. Current quantum computers are noisy, which limits the computation achievable by them. In the absence of error correction and fault-tolerance, other methods to suppress \cite{viola1999dynamical, nation2023suppressing} and mitigate \cite{temme2017error, van2022model, nation2021scalable, van2023probabilistic, kwon2020hybrid} the noise in the system have been studied widely. Several studies exploit one or more of these methods to perform reliable computation involving hundreds of qubits \cite{kim2023evidence, shtanko2023uncovering}.

Apart from these, circuit cutting is a promising approach that has been proven to reduce the noise in the system. A given quantum  circuit is partitioned into multiple subcircuits, which are executed independently and then by using classical postprocessing  the uncut output distribution is retrieved; it was proposed primarily as a method to execute larger circuits on smaller devices \cite{peng2020simulating}. Since then, multiple studies \cite{saleem2021divide, basu2022qer, majumdar2022error} establish their capability to reduce the noise in the system as each of the subcircuits involves fewer qubits and/or gates. In \cite{khare2023parallelizing}, the authors obtained a more accurate estimation of ground state energy of the nearest neighbour Hamiltonian by leveraging circuit cutting and computing each subcircuit on the least-busy device.  In \cite{Chatterjee_2022}, the authors provided the framework of a scheduler to assign circuits to multiple hardware to minimize the overall execution time but neither considered the noise profile of the hardware. In this study, we propose a noise and time-aware scheduler that leverages circuit cutting and then schedules the subcircuits to multiple hardware to maximize the overall fidelity while restricting the execution time on each hardware to a pre-specified value $\tau$.

\textit{Motivation:} A user would ideally want to (i) reduce the effect of noise on the quantum circuit, and (ii) execute their quantum circuit on the least noisy hardware. These two requirements are not independent, but we leverage circuit cutting and perform noise-aware scheduling respectively. Circuit cutting is known to lower the effective noise on the system, but it also leads to an increased number of subcircuit instances to be executed (refer to Sec.~\ref{subsec:cutting}). For this study, we assume that an establishment has dedicated access to a set of hardware. However, a user in that establishment may not have the necessary allocation on the hardware of their choice. This may be due to limited access for that user, or due to the access time being shared among multiple users. In our approach, we tackle the challenge of enhancing fidelity while reducing execution time by combining circuit cutting and selecting the best available hardware for execution of each subcircuit.

\textit{Main contribution:} Achieving the optimal trade-off between noise and execution time is a complex task since these two requirements are often orthogonal to each other. In order to minimize the noise, we would want to execute all the subcircuits on the least noisy device available, leading to an increased execution time. Conversely, to minimize the execution time, we can opt to distribute the subcircuits across all available devices without considering their individual noise profiles, leading to low fidelity. In order to address this optimization challenge, we design an integer linear program (ILP) that seeks to maximize the fidelity while conforming to a fixed maximum allowable execution time $\tau$ for each hardware. The uncut probability distribution is obtained through classical postprocessing over the outcomes of the individual subcircuits. The results obtained through our Noise and Time Aware Distributed Scheduler ($NoTaDS$) demonstrate significantly better fidelity for different benchmark circuits, compared to the scenario where the uncut circuit was executed on the least noisy device. Moreover, if the number of subcircuits is not more than the number of hardware, then, with certain restrictions on the maximum execution time for each hardware are maintained, we provide a polynomial time graph theoretical scheduler. Its  results are equivalent to the ILP scheduler for these cases. Our method represents an initial step towards noise and time-minimized distributed quantum computing, showcasing promising outcomes in improving the performance of quantum computing in real-world applications.
Note that for this study we have assumed dedicated access to quantum hardware for an establishment. Therefore, the issue of queuing delay does not arise in our problem setting. 

The rest of the paper is organized as follows: In Section \ref{2}, we present the basic concepts of quantum circuit fragmentation, circuit placement, and how to select {\it good} qubits. We discuss  about hardware schedule for subcircuits in Section \ref{3}. We propose the time and noise-optimized distributed scheduler in Section \ref{sec:ilp}. Section \ref{sec:experiments} reports the experimental results of the proposed methodology. In Section~\ref{sec:matching}, we provide a polynomial time algorithm for scheduling under certain restrictions on the number of subcircuits and the execution time for each hardware. Section \ref{sec:discussion} has the conclusion.

\section{Background}\label{2}

Noise is arguably the primary hindrance to the scalability and applicability of quantum computers for problems of interest. While error correction is the primary goal to achieve in the long run, current quantum devices do not have the necessary qubit count and/or low enough noise profile for it. Therefore, methods to suppress \cite{viola1999dynamical, nation2023suppressing} or mitigate \cite{temme2017error, nation2021scalable, van2022model, van2023probabilistic, kim2023evidence} the effect of noise are widely studied for near-term quantum devices. In this study, we have made use of two error suppression methods -- circuit cutting \cite{peng2020simulating, tang2021cutqc} and selection of \emph{good} qubits \cite{nation2023suppressing} for mapping the circuit onto the hardware in order to cope with the noise in the system. We make use of both of these methods to propose an optimized scheduling of circuits on the available hardware which aims to minimize both the noise and the overall execution time. In the next two subsections, we briefly discuss the two mentioned methods.

\subsection{Circuit cutting}
\label{subsec:cutting}
With limitations in the size of current hardware, methods to partition a circuit into multiple smaller subcircuits have been studied widely. These methods include splitting the problem itself to execute multiple smaller subcircuits (e.g. entanglement forging \cite{eddins2022doubling}), cutting the circuit between two gates to create multiple tomographic instances of smaller subcircuits (called wire cutting \cite{peng2020simulating}), or replacing two-qubit gates by multiple instances of single qubit operation and feedforward classical communication (called gate cutting \cite{mitarai2021constructing}). In this manuscript, we focus on wire cutting only, and use the term \emph{circuit cutting} to imply wire cutting.

Given a circuit $\Phi$, let the expectation value of an observable $A$ be denoted as $\Phi(A)$. Now for any observable $A$, it is possible to write \cite{tang2021cutqc}
\begin{center}
    $A = \frac{Tr\{A.I\}I + Tr\{A.X\}X + Tr\{A.Y\}Y + Tr\{A.Z\}Z}{2}$
\end{center}

where $I, X, Y, Z$ are the Pauli operators \cite{nielsen2010quantum}. In other words, $\Phi(A) = \frac{1}{2}\sum_{P \in \{I,X,Y,Z\}}c_P\Phi_P(A)$, where $\Phi_P(A) = Tr\{AP\}\rho_P$. Here $\rho_P$ denotes the eigenstates of the Pauli operator $P$ and $c_P$ denotes the eigenvalue. Note that the mathematical expression $Tr\{AP\}\rho_P$ takes instances of both subcircuits into account where the former is measured in basis $P$ and the latter is prepared in the state $\rho_P$. Since there are two eigenstates corresponding to each Pauli operator, this method results in four subcircuit instances for measurement basis and eight for preparation state. The uncut expectation value (or probability distribution) is obtained via classical postprocessing.

In \cite{tang2021cutqc} the authors showed that the previous representation of the observable $A$ is tomographically over-complete; It is possible to have a more succinct representation of $\Phi(A) = \sum_i Tr\{AO_i\}\rho_i$, where $O_i \in \{X,Y,Z\}$ and $\rho_i \in \{\ket{0},\ket{1},\ket{+},\ket{+i}\}$. These two sets $O_i$ and $\rho_i$ are tomographically complete and hence denote the minimum number of subcircuits necessary. Here, there are three subcircuit instances for measurement basis and four for preparation state. A general drawback of cutting is that the classical postprocessing time scales exponentially in the number of cuts when the full probability distribution needs to be reconstructed. Therefore, this method is suitable only for circuits that can be split into disjoint subcircuits using a small (ideally constant) number of cuts only.

\begin{figure}[htb]
    \centering
        \includegraphics[scale=0.2]{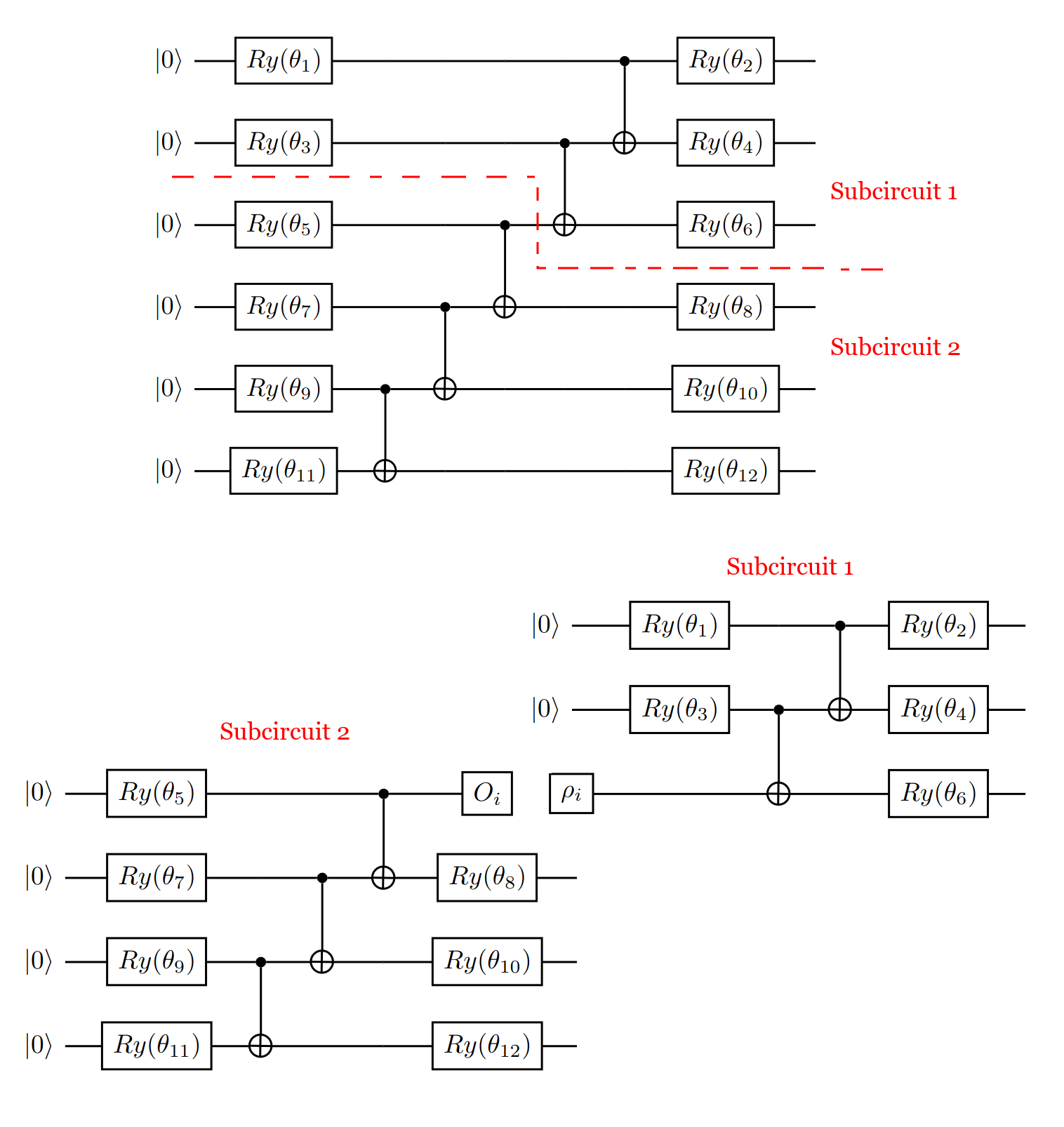}

    \caption{Circuit cutting for a 6-qubit RealAmplitudes ansatz, with single repetition and reverse-linear entanglement \cite{kandala2017hardware}, into two subcircuits. }
    \label{fig:realamplitude}
\end{figure}

Let us consider a RealAmplitudes \cite{kandala2017hardware} circuit with linear reverse entanglement with a single repetition. An $n$-qubit RealAmplitudes circuit consists of $n-1$ CNOT gates and two layers of $R_y$ gates, resulting in $2n$ parameters. Fig.~\ref{fig:realamplitude} shows circuit-cutting of a 6-qubit RealAmplitudes circuit resulting in two subcircuits. The cut is denoted by the dotted red line. Here $\rho_i$ and $O_i$ have similar meaning as discussed above. Therefore, there are three variants of the first subcircuit for $O_i = X, Y, Z$, and four of the second for $\rho_i = \ket{0},\ket{1},\ket{+},\ket{+i}$.

As each subcircuit has a lower number of qubits and/or gates, the noise on each subcircuit is expected to be lower. Hence circuit cutting is often used as a method to reduce the noise in the system \cite{tang2021cutqc, basu2022qer, majumdar2022error, khare2023parallelizing}. In other words, the motivation for circuit cutting is not only the ability to run bigger circuits on smaller hardware but also to lower the noise in the system at the cost of some classical post-processing.

\subsection{Circuit placement and selection of {\it good} qubits}
\label{subsec:mapomatic}

\begin{figure}[htb]
    \centering
    \includegraphics[scale=0.35]{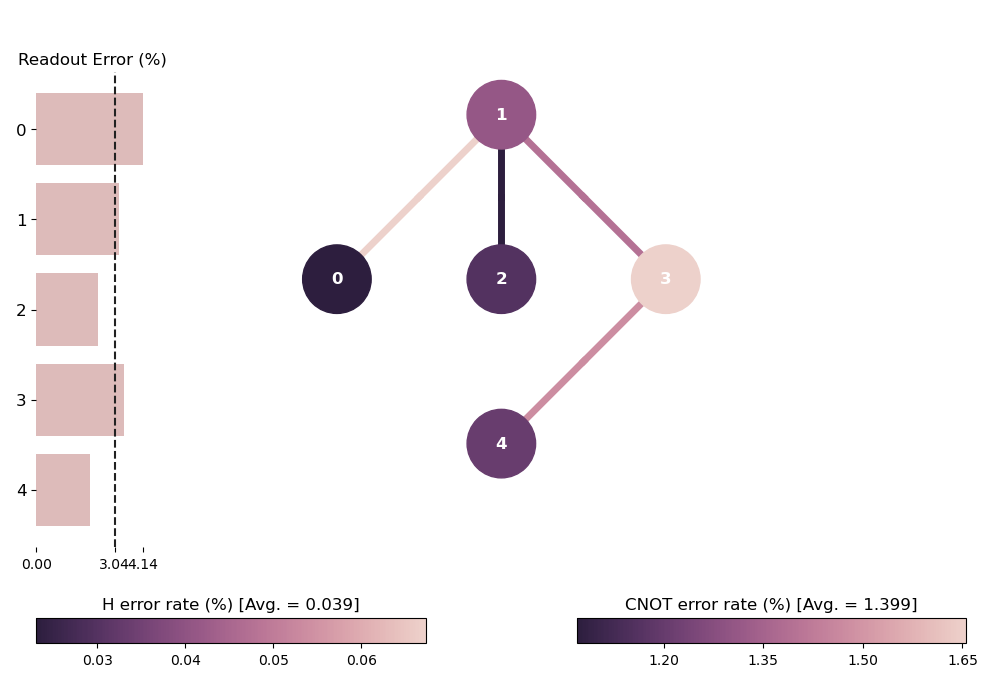}
    \caption{The coupling map and error distribution of a 5 qubit IBM Quantum device $Belem$.}
    \label{fig:belem}
\end{figure}

In current quantum devices, a two-qubit operation is possible only between nearest neighbours. For example, Fig.~\ref{fig:belem} shows the coupling map of a 5-qubit IBM Quantum device. Here a two-qubit operation is possible between qubits 0 and 1, but not between 0 and 2 since the latter are not neighbours. In order to perform a two-qubit operation between qubits 0 and 2, they must be adjacent to each other using SWAP gates. A general requirement of placement and scheduling algorithms \cite{10035992, tan2020optimality, li2020qubit, sivarajah2020t, Qiskit, zulehner2017one, kole2019improved, childs2019circuit, murali2019noise} is to minimize the number of SWAP gates.

Although the aim of the placement is to minimize the number of SWAP gates, Fig.~\ref{fig:belem} clearly shows that the noise profile of all the qubits is not the same. Therefore, it is important to try to involve the less noisy, or \emph{good}, qubits from the hardware for placement. However, selecting \emph{good} qubits for placement may lead to increased SWAP gates if the \emph{good} qubits are not adjacent. Therefore, minimization of SWAP gates and selection of \emph{good} qubits can often be contradictory requirements in placement.

\begin{figure}[htb]
    \centering
    \includegraphics[scale=0.2]{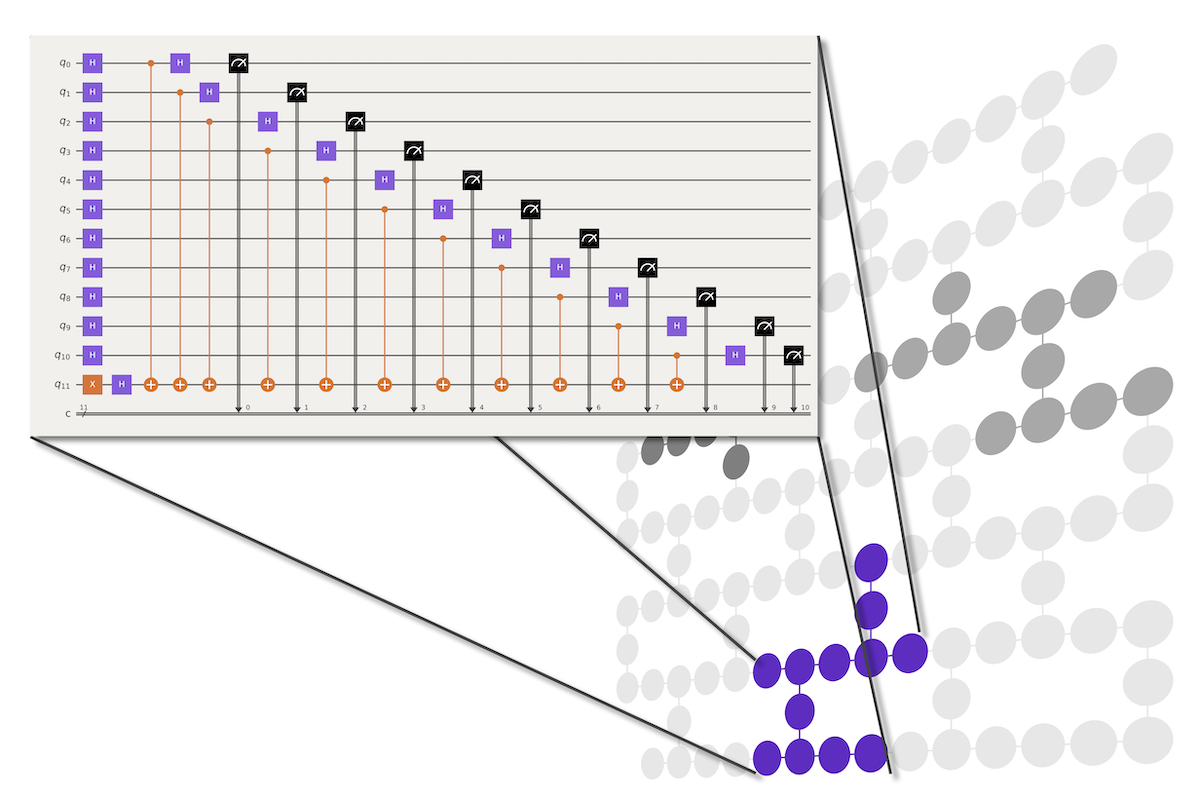}
    \caption{An example of \emph{mapomatic} \cite{nation2021scalable} to find the best placement of a circuit on hardware. This figure is obtained from the GitHub repository of mapomatic (https://github.com/Qiskit-Partners/mapomatic).}
    \label{fig:mapomatic}
\end{figure}

In \cite{nation2023suppressing}, the authors proposed a two-step solution for this. In the first step, also known as \emph{transpilation}, the placement algorithm focuses on minimizing the number of SWAP gates without considering the noise profile of the hardware. As a second step, a list of isomorphisms of the transpiled circuit graph on the hardware graph is generated (refer to Fig.~\ref{fig:mapomatic}). Each of these isomorphisms is also called \emph{layout}. Finally, the noise profile of each layout is calculated from the calibration data of the hardware to assign a score $Q$ which is an indicator of the quality of the layout. The layout having the lowest score, which corresponds to the best quality, is selected. This entire process has been named \emph{mapomatic} by the authors. In this paper, we use \emph{mapomatic} for the selection of the best qubit placement for a given circuit.

For a set of hardware, Table~\ref{table:mapomatic} shows the least mapomatic score and the corresponding layout for placement of a 6-qubit RealAmplitudes circuit  (Fig.~\ref{fig:realamplitude}). In other words, each layout shown in the table implies that both the number of SWAP gates and the noise will be minimized if the circuit is placed on those qubits of the hardware. A layout is generally represented as an array $l$, where $l[k]$ denotes the qubit of the hardware on which the $k-th$ qubit of the circuit is mapped. For example, from Table~\ref{table:mapomatic}, in IBMQ Hanoi, the qubits 0, 1, 2, 3, 4, and 5 of the 6-qubit Real Amplitudes circuit are respectively mapped to physical qubits 0, 1, 2, 4, 7 and 6. From this table, we get that IBMQ Kolkata is the best hardware with layout [22, 25, 26, 24, 23, 21] to execute the 6-qubit Real Amplitudes circuit.

\begin{table}[htb]
\centering
\caption{$Mapomatic$ score for the best layout of the 6-qubit RealAmplitudes circuit (Fig: \ref{fig:realamplitude}) corresponding to each of the available hardware}
\adjustbox{scale=0.8}{
\begin{tabular}{|c|c|c|c|}
\hline
  Backend & \# Qubits  & Corresponding layout & Mapomatic score\\
 \hline
IBMQ Hanoi & 27 & [0, 1, 2, 4, 7, 6]  & 0.099\\
 IBMQ Mumbai &  27 & [6, 7, 4, 10, 12, 13] & 0.183 \\
IBMQ Cairo & 27 & [13, 12, 10, 15, 18, 17]  & 0.105\\
IBMQ Kolkata &27  & [22, 25, 26, 24, 23, 21] & 0.084 \\
IBMQ Guadalupe & 16 & [15, 12, 13, 10, 7, 6] &  0.142 \\
IBMQ Lagos & 7 & [0, 1, 2, 3, 5, 6] & 0.093  \\
IBMQ Nairobi &7  & [0, 1, 2, 3, 5, 4] & 0.193 \\
 
 \hline

 \end{tabular}
 }

 \label{table:mapomatic}
 \end{table}

In the next section, we leverage circuit cutting and \emph{mapomatic} to maximize the fidelity of a (or a set of) circuits given a set of hardware while minimizing the overall execution time.

\section{Hardware Schedule for Subcircuits}\label{3}

In this paper, we study the problem of scheduling jobs on different hardware with a focus on maximizing the fidelity and minimizing the execution time. First, we want to emphasize that in this paper we consider circuit-cutting primarily as a method to suppress the effect of noise. It has been shown in multiple studies that circuit cutting itself can lower noise in the system \cite{basu2022qer, saleem2021divide, ayral2021quantum, majumdar2022error}. Therefore, we shall resolve to circuit cutting even if the circuit is small enough to be executed on the hardware. This method allows us to improve fidelity as well as use parallel scheduling of the subcircuits obtained after cutting to multiple hardware, thus lowering the execution time \cite{Chatterjee_2022}.

Consider a list of hardware $H$ and a list of circuits (or subcircuits) $C$. For a (sub) circuit $i \in C$, let $l_{ij}$ be the optimum (least noisy) layout on a hardware $j \in H$. Ideally,  each (sub) circuit can be assigned to the best layout corresponding to it, in terms of noise. However, cutting the circuit increases the number of executable circuits by creating multiple instances for each subcircuit (refer to Fig.~\ref{fig:realamplitude}). Therefore, the trade-off for error suppression using circuit cutting is the increased execution time to execute all the subcircuit instances.

If the number of hardware available is at least as many as the number of subcircuit instances, then a polynomial time algorithm for finding a minimum weight maximum matching in the bipartite graph having an edge between a subcircuit and a hardware with a weight (say, the $mapomatic$ score) can provide the required assignment. This also comes with an inherent assumption that the maximum execution time for each hardware can accommodate no more than one subcircuit instance. However, if the number of subcircuit instances are more than that of available hardware, or the allowed execution time for each hardware can accommodate more than one subcircuit instance, then job scheduling has to be performed. 

Of these multiple subcircuits, most of them may conform to the least noisy layout on the same hardware. Therefore, the best assignment may lead to \emph{sequential} execution of a large number of subcircuits, leading to a large execution time. A user often has limitations on the execution time on a particular hardware, barring this sequential approach. On the other hand, the execution time can be minimized if we opt for as much parallelization as possible, i.e., equally distribute the (sub) circuits to all the available hardware without considering the noise profile.

In this study, we delve into finding the optimum scheduling of the (sub) circuits on the available hardware, when an upper limit on the execution time for each hardware is imposed, such that the overall fidelity is maximized. The problem statement can be formally stated as follows:

\begin{definition}
    Given a list of circuits $C$, a list of hardware $H$ and the corresponding execution time limit $\tau_j$ for  $j \in H$, find an assignment $X_{ij}$ $\forall$ $i \in C$ such that the fidelity of the circuits are maximized and the execution time $t_j \leq \tau_j$ $\forall$ $j \in H$.
\end{definition}

This problem, in general, is not known to have a polynomial time solution. However, if (i) the number of subcircuits is at most as many as the number of hardware, and (ii) the maximum allowable time $\tau_j$ for each $j \in H$ can accommodate only one subcircuit. We first propose an integer linear programming (ILP) approach to the scheduling problem in general in Sec.~\ref{sec:ilp}, elaborated in Fig.~\ref{fig:flowchart}. Later, in Sec.~\ref{sec:matching} we show a polynomial time solution to this problem under the above mentioned restrictions.

\section{Proposed framework}
\label{sec:ilp}
\begin{figure}[htb]
    \centering
    \includegraphics[scale=0.38]{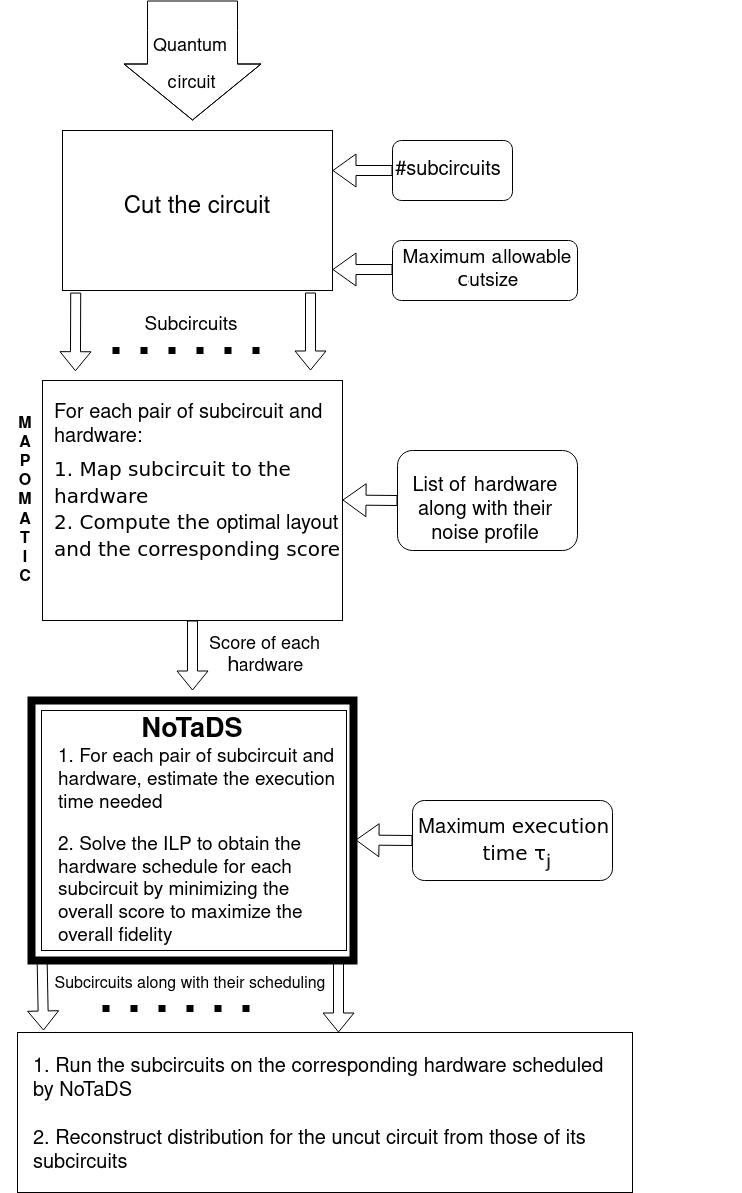}
    \caption{A flowchart of our proposed noise and time optimized scheduler $NoTaDS$, including circuit cutting, scoring of (circuit, hardware) pair, noise and time optimized scheduling, and final reconstruction of the entire probability distribution from those of the subcircuits.}
    \label{fig:flowchart}
\end{figure}
We start with the premise where a list of circuits $C$ and a list of hardware $H$ are provided. The list of hardware can either be provided by the user or can be determined from their credential for a particular vendor. For each $c \in C$, we first fragment it into $k$ subcircuits via circuit cutting. Note that as stated before, we resort to cutting all the circuits, irrespective of whether these can be executed on a single hardware, in order to reduce the noise, and thus improve the fidelity. Henceforth, $C$ denotes the set of all subcircuits obtained via circuit-cutting, and $i \in C$ implies a subcircuit.

The steps in the flowchart of Fig.~\ref{fig:flowchart} are described in the following subsections.

\subsection{Selection of appropriate hardware}

As stated before, let $C$ be the set of all subcircuits. Naturally, tagging is required to keep track of which subcircuit corresponds to which circuit for classical recombination over the cuts (refer to Sec.~\ref{subsec:cutting}) which follows later. First, for each subcircuit $i \in C$, the set of hardware $H_i \in H$ is determined such that for all $j \in H_i$ the number of qubits in $j$ is at least as big as the number of qubits in the subcircuit $i$. If $H_i = \{\}$ for any subcircuit $i$, then $i$ needs to be partitioned again such that at least one hardware can accommodate each subcircuit.

At the end of this step, we obtain a list of feasible hardware $H_i$ for each subcircuit $i$.

\subsection{Scoring each hardware as per noise profile}

Next, we use mapomatic \cite{treinish2022mapomatic} for each $i \in C$ and $j \in H_i$. The action of mapomatic here can be considered as a function
\begin{center}
    $\mathcal{F}: \{j, i\} \rightarrow \{l_{ij},Q_{ij}\}$
\end{center}

where $l_{ij}$ is the optimum layout and $Q_{ij}$ is the mapomatic score for this set of circuits and layout. In other words, given a hardware $j$ and a circuit $i$, mapomatic returns a list of physical qubits $l$, which is the initial layout for the placement of $i$ on $j$, and a score $Q_{ij}$, which is an indicator of the noise. For each (subcircuit-hardware) pair $(i,j)$, we obtain a set of such score $Q_{ij}$. Therefore, for each circuit $i$, this step produces a list of hardware $j \in H_{i}$ ordered by the score $Q_{ij}$. In the usual scoring technique of mapomatic, a lower score implies a lower noise profile. Therefore, a hardware $h_1$ is better than $h_2$ for a circuit $i$ if $Q_{ih_{1}} < Q_{ih_{2}}$. However, one may define their own custom scoring technique which may imply the opposite.
At the end of this step, we obtain an ordering among the list of feasible hardware for each subcircuit in terms of their noise profile.

\subsection{Noise and Time Aware Distributed Scheduler }

We propose an integer linear program (ILP) to schedule each subcircuit from $C$ to quantum hardware such that the fidelity is maximized while conforming to the upper bound on the execution time for each hardware. Note that the list of feasible hardware and their ordering as per mapomatic score may vary from circuit to circuit. Therefore, the optimization needs to take into account this variation for each subcircuit.

After cutting, each subcircuit corresponds to multiple subcircuit instances (refer to Sec.~\ref{subsec:cutting}). The number of instances of a subcircuit $i$ depends on the number of preparation qubits $\rho_i$ and the number of measurement qubits $O_i$. We associate a value $\eta_i$ with each subcircuit $i$ such that
$$
    \eta_i = \begin{cases}
        1 & \text{if all instances are scheduled individually} \\
        \nu (\rho_i, O_i) & \text{otherwise}
    \end{cases}
$$

where $\nu (\rho_i, O_i)$ denotes the total number of subcircuit instances for subcircuit $i$. In Sec.~\ref{sec:experiments} we discuss their advantages and disadvantages.

Next, we define the variables, constraints, and the objective function for the ILP.

\begin{enumerate}
    \item Variables: We associate a variable $X_{ij}$ for each subcircuit $i \in C$ and hardware $j \in H_i$ such that
$$
X_{ij} = \begin{cases}
    1 & \text{if subcircuit $i$ is scheduled to hardware $j$} \\
    0 & \text{otherwise}.
\end{cases}
$$

In other words, $X_{ij}$ acts as a decision variable for the scheduling. Moreover, a score variable $Q_{ij}$ is associated with each $X_{ij}$ which indicates the mapomatic score when subcircuit $i$ is placed on hardware $j$.


\item Constraints:\label{constraint} Next, we fix the constraints for the ILP. 

\begin{enumerate}
    \item The first requirement is that every subcircuit $i$ is assigned to some hardware. Formally, this constraint can be represented as
    \begin{equation}
    \label{eq:constraint1}
    \sum_{j \in H_i} X_{ij} = 1.
    \end{equation}
    Note that this constraint should hold for all subcircuits $i \in C$, and therefore, Constraint~\ref{eq:constraint1} essentially results in $|C|$ constraints.
    
    \item As mentioned above, there is a time restriction for each hardware for a user. We associate a maximum execution time $\tau_j$ for each $j \in H$. The value of $\tau_j$ can be provided by the user or can be determined from the user's access plan. Let $t_{i}$ denote the execution time for each subcircuit $i \in C$. Then, the total execution time of all the subcircuits scheduled to a particular hardware $j$ should not exceed $\tau_j$. Formally, this is represented as
    \begin{equation}
        \label{eq:constraint3}
        \sum_{i \in C} \eta_i \cdot t_{i} \cdot X_{ij} \leq \tau_j.
    \end{equation}

Note that while the summation of this constraint goes over the entire set of subcircuits, the indicator variable $X_{ij}$ ensures that the time for a particular subcircuit is added to the execution time only if it is scheduled to that hardware. This constraint holds for each hardware, and therefore Constraint~\ref{eq:constraint3} essentially results in $|H|$ constraints.

\end{enumerate}


\item Objective Function: The objective of this optimization problem is to maximize the overall fidelity, which translates to minimizing the overall score $Q$. Therefore, the objective function for this is defined as

\begin{equation}
    \label{eq:objective}
    \min \sum_{i \in C, j \in H} X_{ij} \cdot Q_{ij} \cdot A_i
\end{equation}

Here, $0 < A_i \leq 1$ is the area of circuit $i$ normalized over all available circuits. We define the area of a circuit as the product of the number of qubits and the 2-qubit depth. In other words, higher the area of a circuit, more amenable it is to noise. The inclusion of this term thus ensures that circuits with a higher area are placed in layouts with lower mapomatic scores to ensure the quality of the outcome. For example, Fig. \ref{fig:area} shows a possible cutting of a 5-qubit circuit into two subcircuits. Both the subcircuits consist of 3 qubits, but the 2-qubit depth of the first subcircuit is 7 whereas that of the second subcircuit is 2. The first subcircuit which has more depth is more prone to noise, and hence should be placed in a better layout.
 \begin{figure}[!h]
    \centering
    \includegraphics[scale=0.2]{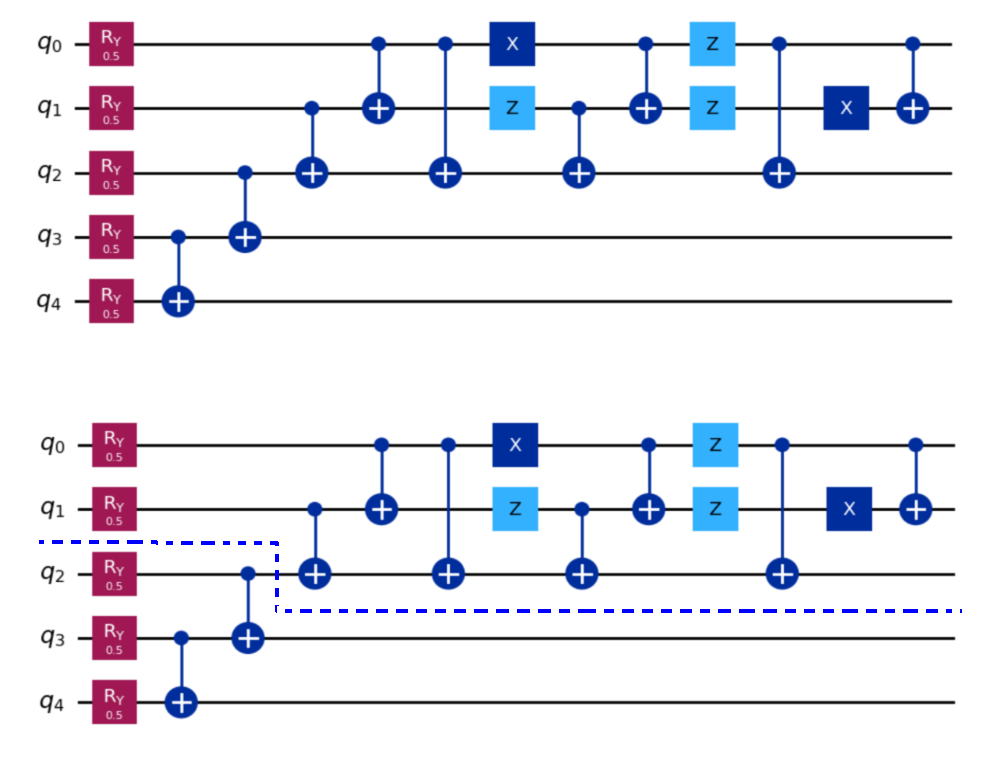}
    \caption{Cutting with unequal depth : to understand why area is an important element in the objective function}
    \label{fig:area}
\end{figure}

The whole ILP formulation thus, is
\begin{eqnarray*}
    \min & & \sum_{i \in C, j \in H} X_{ij} \cdot Q_{ij} \cdot A_i\\
    \text{subject to} & & \text{constraints (a) - (b)}\\ 
    & & X_{ij} \in \{0,1\}.
\end{eqnarray*}
\end{enumerate}

\textit{Note on  linear objective function}
The objective function in Eq.~\ref{eq:objective} is linear in $Q_{ij}$. It’s worth questioning if a linear function is adequate when minimizing over various subcircuits and their instances. While $mapomatic$ scores reflect the hardware noise profile, a shallow subcircuit on a noisy layout might outperform a deep circuit on a less noisy layout. For subcircuits that are approximately equal in qubit count and depth, the hardware layout's $mapomatic$ score mainly reflects its noise profile, making a linear objective function sufficient. However, for subcircuits with significant differences in width or depth, both circuit size and noise affect the $mapomatic$ score, potentially requiring a non-linear objective function for optimal scheduling. Given that previous studies suggest optimal performance with balanced cutting \cite{basu2022qer}, this study focuses on balanced subcircuits, thus a linear objective function is appropriate.


\section{Experimental Results}
\label{sec:experiments}

In this section, we present the experimental results of our $NoTaDS$ scheduler for different types of circuits. We have used the Circuit-knitting-toolbox \cite{circuit-knitting-toolbox} for circuit cutting and reconstruction, and CPLEX Optimization Studio \cite{cplex} to solve the ILP of Sec.~\ref{sec:ilp}. Table~\ref{table:BACKEND} shows the set of quantum hardware used for our experiments and their noise profile. Some of the parameters for the noise profile include the probability of faulty gates and measurement, and the rate of spontaneous decay of a qubit, characterized by $T_1$ and $T_2$. The noise profile of the hardware varies with time. The values for each type of error in the table are the average over all the qubits in that hardware. Moreover, the readout error probability for each qubit is the average of $p(0|1)$ and $p(1|0)$ where $p(i|j)$ denotes the probability of measuring $i$ when the outcome was originally $j$, $i, j \in \{0, 1\}$.

\begin{table*}[htb!]
\centering
\caption{Number of qubits and noise profile of the hardware used in our experiments}
\begin{tabular}{ |c|c|c|c|c|c|c|}
\hline
 
Hardware & \multirow{2}{*}{\# Qubits} & 2-qubit gate & 1-qubit gate & \multirow{2}{*}{$T_1$ ($\mu s$)} & \multirow{2}{*}{$T_2$ ($\mu s$)} & Readout error\\
name & & error probability & error probability & & & probability \\ 
 \hline
IBMQ Hanoi & 27 & $8.3 \times 10^{-3}$ & $2.1 \times 10^{-4}$  & 156.69&  137.7 & $10^{-2}$\\
IBMQ Mumbai & 27 &  $7.5 \times 10^{-3}$ & $2.5 \times 10^{-4}$ & 118.01 & 161.97 & $1.8 \times 10^{-2}$\\
IBMQ Cairo & 27 & $9.4 \times 10^{-3}$ & $2.2 \times 10^{-4}$ & 94.62 &  116.42 & $1.3 \times 10^{-2}$\\
IBMQ Kolkata & 27 & $8.7 \times 10^{-3}$ & $2 \times 10^{-4}$ & 117.42 & 92.97& $1.2 \times 10^{-2}$\\
IBMQ Guadalupe & 16 & $9.74 \times 10^{-3}$ & $2.64 \times 10^{-4}$ & 86.72 & 118.73 & $1.64 \times 10^{-2}$\\
IBMQ Lagos & 7 & $7.2 \times 10^{-3}$ & $2 \times 10^{-4}$ & 112.51 & 84.42 & $1.4 \times 10^{-2}$\\
IBMQ Nairobi &7  &  $8.7 \times 10^{-3}$ &  $ 3.5 \times 10^{-4}$ & 114.75 & 71.42 & $ 2.7 \times 10^{-2}$ \\
IBMQ Jakarta & 7 & $7.3 \times 10^{-3}$ &  $ 1.03 \times 10^{-4}$ & 136.95 & 38.99 & $2.09 \times 10^{-2}$\\
IBMQ Manila &5 & $7.7 \times 10^{-3}$ &  $ 2.46 \times 10^{-4}$  &  141.15&  56.53& $2.2 \times 10^{-2}$ \\
IBMQ Lima & 5 & $9.58 \times 10^{-3}$ &  $ 3.76 \times 10^{-4}$ & 98.68 & 115.32 & $2.41 \times 10^{-2}$ \\
IBMQ Belem & 5 & $8.89 \times 10^{-3}$ &  $ 3.88 \times 10^{-4}$  & 101.42  & 98.85 & $2.39 \times 10^{-2}$ \\
IBMQ Quito & 5 & $7.9 \times 10^{-3}$ &  $ 2.88 \times 10^{-4}$  & 96.83  & 104.39 & $4.15 \times 10^{-2}$ \\
 
 \hline

 \end{tabular}

 \label{table:BACKEND}
 \end{table*}

In the following subsections, first, we discuss the selection of the maximum execution time $\tau$ for each hardware, and then we discuss our method for estimating the execution time of a subcircuit. Finally show the fidelity obtained by our scheduling method for a range of quantum circuits.

\subsection{Criteria for maximum execution time $\tau$}
\label{subsec:tau}

In Sec.~\ref{constraint}, we defined the maximum execution time for a hardware $j \in H$ as $\tau_j$. There are $\eta_i$ instances for each subcircuit $i \in C$. Let $t^{(i)}$ be the execution time for subcircuit $i$. The maximum execution time required for a particular hardware, $\tau_{max}$, is when all the instances of all the subcircuits are executed sequentially on one particular hardware. Then
\begin{center}
    $\tau_{max} = \sum_{i \in C} \eta_i \cdot t^{(i)}$.
\end{center}

Note that allowing any excess time to $\tau_{max}$ does not change the scheduling and execution time. Therefore, we stick to equality instead of $\geq$.

The minimum time $\tau_{min}$ that each hardware should allow should be such that at least one subcircuit can be executed. If there are any hardware which does not conform to this requirement, that one can simply be removed from the list of all available hardware. Here, we want to mention once more that one subcircuit $i \in C$ consists of $\eta_i$ subcircuit \emph{instances}. One can choose to schedule each instance or each subcircuit. We tested the former, which resulted in a drop of fidelity by $\sim 9\%$ over the latter. This is obvious since the different instances are equivalent to a tomography of the subcircuit \cite{perlin2021quantum, majumdar2022error}. Therefore, running each instance of the subcircuit on different hardware (i.e., different noise models) leads to an inaccurate tomography, which makes the reconstruction fallible. Therefore, for this study, we stick to the scheduling of subcircuits, and not the instances. There may be scenarios where scheduling the instances in an intelligent way may lead to a lower decrease in the fidelity -- we postpone that for future studies.

For our experimental settings, the number of hardware is always chosen to be greater than the number of subcircuits. Therefore,
\begin{center}
    $\tau_{min} = \max_{i \in C} \eta_i \cdot t^{(i)}$.
\end{center}

In all our experiments, we fix $\tau_j = \tau_{min}$ $\forall$ $j \in H$. Later in Sec.~\ref{sec:change_fid} we show the change in fidelity and execution time if we allow $\tau_j > \tau_{min}$. 

Calculating the values of $\tau_{max}$ and $\tau_{min}$ require the execution time $t^{(i)}$ for all the subcircuits $i \in C$. In the following subsection, we discuss the method we used to fix the value of $t^{(i)}$ $\forall$ $i \in C$ in our experiment.

\subsection{Estimation of the execution time of a circuit}
\label{subsec:t_i}

\begin{figure}[htb]
    \centering
    \includegraphics[scale=0.35]{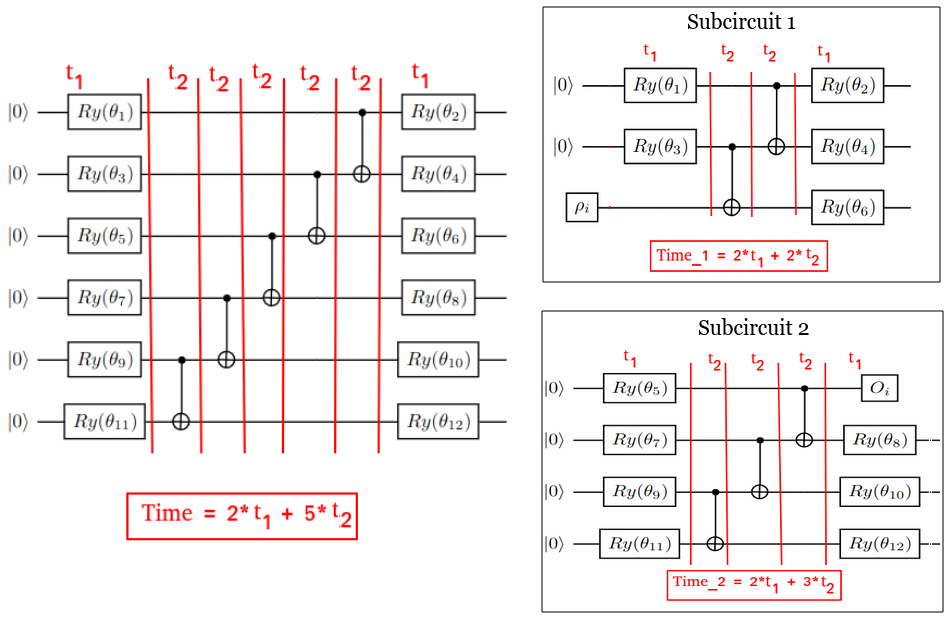}
    \caption{An example of a 6-qubit RealAmplitudes circuit with the level.}
    \label{fig:circ_time}
\end{figure}

There are sophisticated methods for estimating the run-time of a quantum circuit \cite{javadiabhari2023validating}. However, for our experiment, we stick to a simple method of calculating the time of each level of a circuit. We define \emph{level} of a circuit as the timestamp where some gates are executed in parallel. In Fig.~\ref{fig:circ_time} we separate the different levels of the circuit by red lines.


The time duration of each level is determined by the longest gate in that level. Naturally, 2-qubit gates have a much larger execution time than 1-qubit gates. Therefore, if a level contains a 2-qubit gate, then the time duration of that level is $t_2$, which is the execution time of a single 2-qubit gate. Note that it doesn't matter if the level contains multiple 2-qubit gates since they are operated parallelly. On the other hand, if a level contains only single qubit gates then the time duration of that level is $t_1$, which is the execution time of a single 1-qubit gate. Therefore, if a circuit contains $\kappa_1$ levels where only 1-qubit gates are present and $\kappa_2$ levels where 2-qubit gates are also present, then the overall runtime is $\kappa_1 \cdot t_1 + \kappa_2 \cdot t_2$. In Fig.~\ref{fig:circ_time}, in the whole circuit $\kappa_1 = 2 $ and $\kappa_2 = 5$. In the first subcircuit, $\kappa_1 = 2 $ and $\kappa_2 = 2$, and in the second subcircuit $\kappa_1 = 2 $ and $\kappa_2 = 3$.

In current IBM Quantum devices, the execution time of a CNOT gate is $\sim 10 \times$ that of single-qubit gates. For this study, we assume $t_1 = 1$, making $t_2 = 10$. From the abstraction, the execution time of the circuit in Fig.~\ref{fig:circ_time} is $2 \cdot t_1 + 5 \cdot t_2$. This abstract calculation of the execution time keeps the method simple. Since the values of $\tau_{min}$ and $\tau_{max}$ depend on the execution time, if some other method for determining the execution time is used, or if absolute execution times are selected, then the values of $\tau$ will change accordingly without hampering the assignment of the subcircuits on the hardware. Note that if absolute values are used instead of $t_1$, $t_2$, then one should also account for the fact that the absolute values for the execution time of gates are not always the same on different hardware.

\subsection{Experimental results on 6-qubit circuits}

In Table \ref{table: result1} we consider four benchmark circuits having 6 qubits each. These circuits are small enough to be executed on any hardware with $\geq 7$ qubits, and hence distributed scheduling using circuit cutting may not be deemed necessary here. However, Table~\ref{table: result1} shows that distributed scheduling using circuit cutting still helps in the improvement of fidelity. For each of the circuits, we provide its fidelity with the ideal simulation both without any error mitigation and with measurement error mitigation (MEM). For MEM, we have used the default \textit{MThree} mitigation \cite{nation2021scalable} provided in \textit{Qiskit Runtime} by setting the \textit{resilience level} option to 1.

\begin{table*}[htb]
\centering
\caption{Fidelity for 6-qubit circuits by scheduling over the hardware in Table~\ref{table:BACKEND} with and without circuit cutting for no error mitigation (NoMit ) and measurement error mitigation (MEM).}
\begin{tabular}{|c|c|c|c|c|c|c|c|}
\hline
\multirow{3}{*}{Benchmark circuit} & \multirow{3}{*}{\# qubits} & \multirow{3}{*}{Cut size} & \multirow{3}{*}{\# subcircuits} & \multicolumn{4}{c|}{Fidelity} \\
\cline{5-8}
& & & & \multicolumn{2}{c|}{Uncut} & \multicolumn{2}{c|}{Cut} \\
\cline{5-8}
& & & & NoMit  & MEM & NoMit  & MEM \\
\hline
Ripple carry adder \cite{cuccaro2004new}  & 6 & 2 & 2 & 0.759	&0.787 & 0.792 & 0.843 
\\\hline 
RealAmplitudes \cite{kandala2017hardware} & 6 & 1 & 2 & 0.959 & 0.983 & 0.987 & 0.997
\\\hline 
Trotterized \cite{majumdar2022error} & 6 &2 &2 & 0.922 &  0.951 & 0.965 & 0.974 
\\\hline
Bernstein Vazirani \cite{Qiskit} & 6 & 1 & 2 & 0.81 & 0.869  &  0.882 & 0.944
 \\
\hline 
\end{tabular}

\label{table: result1}
\end{table*}

Naturally, MEM improves the fidelity over no-mitigation. However, we observe that distributed scheduling with circuit cutting without any error mitigation outperforms the fidelity of the uncut circuit with MEM. In this experiment, we partitioned the circuit into two subcircuits. We want to emphasize here that (i) as discussed before, $\tau$ for all hardware was fixed to $\tau_{min}$, and (ii) the uncut circuit was always executed on the best hardware and its corresponding layout as per mapomatic. We obtain an average percentage improvement in fidelity for distributed scheduling using circuit cutting over no cutting by $\sim 5.2$ when no mitigation was used, and by $\sim 4.89$ when MEM was used. The average is taken over the four circuits in Table~\ref{table: result1}.

Next, we dive deeper into the exact details of the experiment for the 6-qubit Ripple carry adder circuit \cite{cuccaro2004new}. This is meant to provide an overall idea for recreating the experimental steps for the circuit in Table~\ref{table: result1} and also those in later subsections. The steps follow from the flowchart provided in Fig.~\ref{fig:flowchart}.

\subsubsection{Details for the 6-qubit Ripple carry adder circuit}

\begin{figure}[htb]
	\centering
	       \includegraphics[scale=0.3]{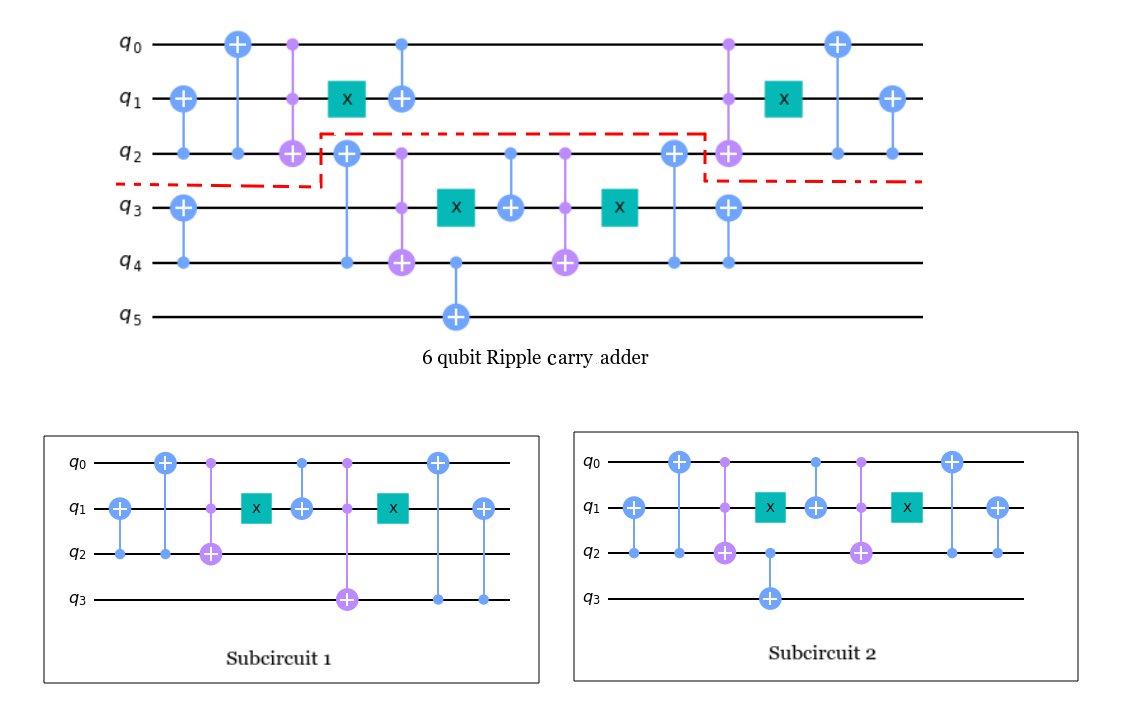}
        \caption{Two subcircuits of 6-qubit Ripple carry adder circuit obtained by using Circuit-knitting-toolbox \cite{circuit-knitting-toolbox}.}
        \label{fig:adder_cut}
\end{figure}

Fig.~\ref{fig:adder_cut} shows the circuit for the 6-qubit ripple carry adder and its two subcircuits obtained using the Circuit-knitting-toolbox \cite{circuit-knitting-toolbox}. After obtaining the subcircuits, we found the hardware big enough to accommodate each of them. In this particular scenario, all the hardware from Table~\ref{table:BACKEND} can accommodate each subcircuit.

Next, we use mapomatic to find the score for each subcircuit against each hardware and its layout. Then we use the optimization in Sec.~\ref{sec:ilp} to schedule the subcircuits to the hardware. Finally, we use mapomatic to find the best hardware and its layout for the uncut circuit as well. In Table~\ref{tab:adder_sched} we show the layout, backend, and the mapomatic score for the uncut circuit and the two subcircuits.

\begin{table}[htb]
    \centering
    \caption{Scheduling details of the 6-qubit ripple carry adder circuit and its two subcircuits obtained after cutting.}
    \adjustbox{scale=0.75}{
    \begin{tabular}{|c|c|c|c|c|}
    \hline
    Circuit & \# qubits & Layout & Scheduled backend & Mapomatic score \\
    \hline
       Uncut & 6 & [26, 25, 24, 23, 21, 18] & IBMQ Kolkata & 0.28 \\
       \hline
       Subcircuit 1 & 4 & [26, 25, 22, 19] & IBMQ Mumbai & 0.13 \\
       \hline
       Subcircuit 2 & 4 & [4, 1, 2, 3] & IBMQ Hanoi & 0.11 \\
       \hline
    \end{tabular}
    }
    \label{tab:adder_sched}
\end{table}

Note that the cut-size to partition the 6-qubit adder circuit into two subcircuits is 2. Therefore, each subcircuit has 4 qubits. We note from Table~\ref{tab:adder_sched} that our scheduler has scheduled the two subcircuits on two different hardware, each of which has a mapomatic score lower than the one where the original circuit has been scheduled. This provides an explanation as to why the fidelity obtained via cutting exceeds the uncut circuit.

\subsection{Experimental results on 10-qubit circuits}

Next, in Table \ref{table: result2} we take a few 10-qubit circuits. These circuits are too big to be executed on 5 or 7-qubit devices but can be executed without cutting on 16 or 27-qubit devices. However, as before, we show that the fidelity can be improved by using our $NoTaDS$ scheduler. Table~\ref{table: result2} shows the fidelity of four 10-qubit circuits with and without measurement error mitigation, where the value of $\tau$ is set to $\tau_{min}$ for all hardware.

\begin{table*}[htb]
\centering
\caption{Fidelity for 10-qubit circuits by scheduling over the hardware in Table~\ref{table:BACKEND} with and without circuit cutting for no error mitigation (NoMit ) and measurement error mitigation (MEM).}
\begin{tabular}{|c|c|c|c|c|c|c|c|}
\hline
\multirow{3}{*}{Benchmark circuit} & \multirow{3}{*}{\# qubits} & \multirow{3}{*}{Cut size} & \multirow{3}{*}{\# subcircuits} & \multicolumn{4}{c|}{Fidelity} \\
\cline{5-8}
& & & & \multicolumn{2}{c|}{Uncut} & \multicolumn{2}{c|}{Cut} \\
\cline{5-8}
& & & & NoMit  & MEM & NoMit  & MEM \\
\hline
Ripple carry adder \cite{cuccaro2004new}  & 10 & 2 & 2 & 0.315	& 0.325 & 0.375 & 0.5138 
\\\hline
Bernstein Vazirani \cite{Qiskit}  & 10 & 1 & 2 & 0.702  & 0.714 & 0.728 & 0.749 \\
\hline
RealAmplitudes \cite{kandala2017hardware} & 10 & 1 & 2 & 0.806 & 0.876 & 0.977 & 0.994
\\\hline 
Trotterized \cite{majumdar2022error} & 10 & 2 & 2 & 0.878 & 0.891 & 0.927 & 0.960
\\\hline 
\end{tabular}

\label{table: result2}
\end{table*}


Once more we observe that the fidelity of the noisy and the ideal circuit obtained without any error mitigation via our scheduling method outperforms (sometimes significantly, e.g., see RealAmplitudes and Trotterized circuits) the fidelity of the uncut circuit with MEM. We obtain an average improvement in fidelity for distributed scheduling using circuit cutting over no cutting by $\sim 12.38\%$ when no mitigation was applied, and by $\sim 21\%$ when MEM was used. The average is taken over the four circuits in Table~\ref{table: result2}.

In the following subsection, we take a deeper dive into the improvement in fidelity with the variation in the number and size of the subcircuits.

\subsection{Variation in fidelity with the number and size of subcircuits}
\begin{figure}[htb]
    \centering
    \includegraphics[scale=0.5]{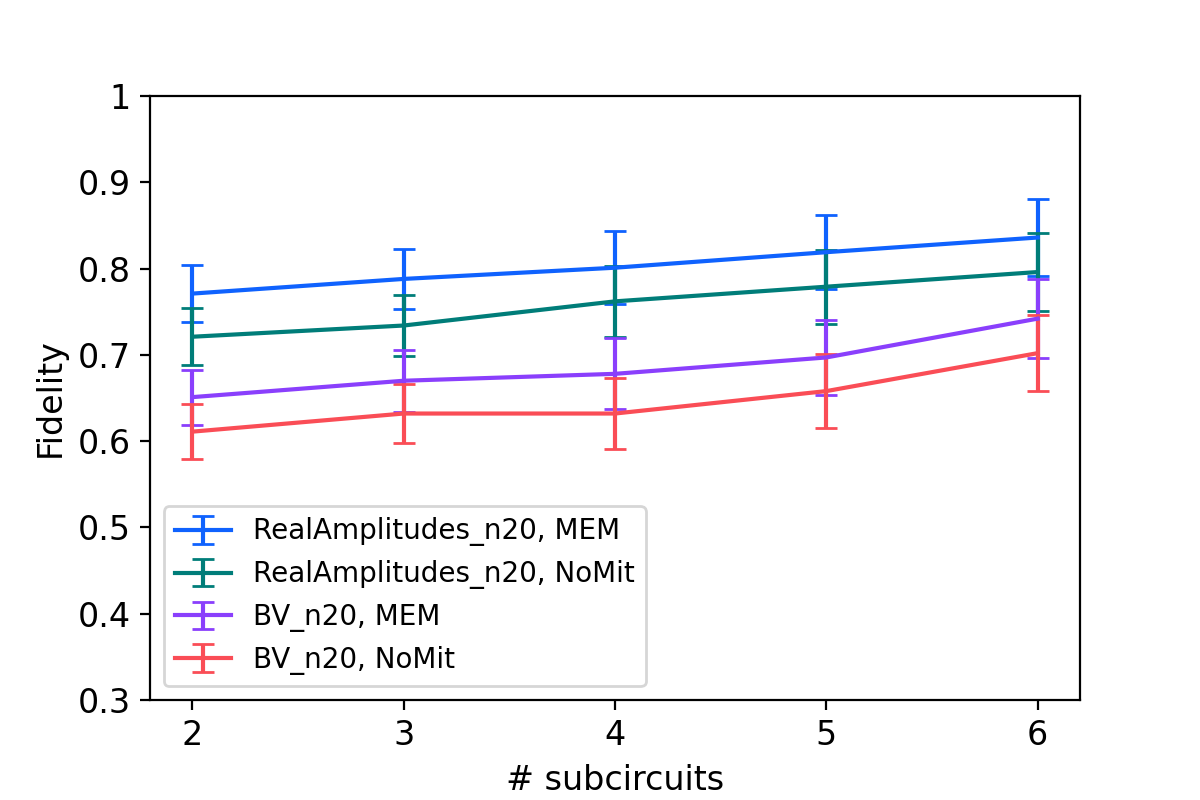}
    \caption{Fidelity obtained by the $NoTaDS$ scheduler with an increasing number of subcircuits for 20-qubit RealAmplitudes and Bernstein Vazirani (BV) circuits.}
    \label{fig:realamp20}
\end{figure}

In Fig.~\ref{fig:realamp20} we plot the fidelity of 20-qubit RealAmplitudes and Bernstein-Vazirani circuits as the number of subcircuits is increased linearly from 2 to 6. In each case, each subcircuit is scheduled using the $NoTaDS$ scheduler, and the fidelity is compared with the ideal outcome. We notice that the fidelity increases linearly with an increasing number of subcircuits.

\begin{figure}[htb]
    \centering
    \includegraphics[scale=0.48]{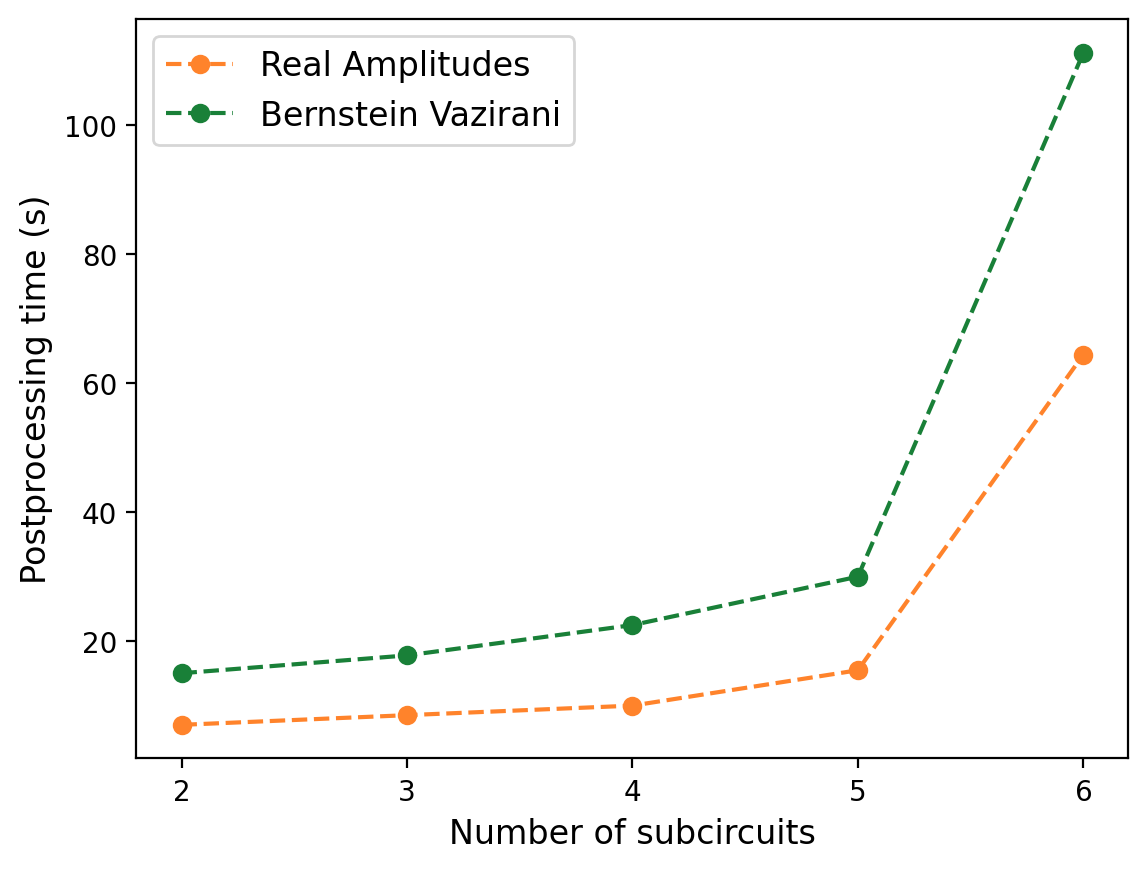}
    \caption{The increase in classical reconstruction time of the full probability distribution from the subcircuits with increasing number of subcircuits.}
    \label{fig:pp}
\end{figure}


We consider cutting 20-qubit RealAmplitudes and Bernestein-Vazirani circuits, where the number of subcircuits varies from 2 to 6. The subcircuits are then scheduled with our proposed $NoTaDS$ scheduler. The result is bootstrapped over 10 trials. We notice a linear improvement in fidelity with the increase in the number of subcircuits. As the number of subcircuits increases, each subcircuit becomes smaller, and hence incurs less contagious noise. Therefore, the fidelity is increased. However, with an increase in the number of subcircuits, the cut-size also increases leading to an exponential increment in the classical postprocessing time for reconstruction of the full probability distribution from the subcircuits \cite{peng2020simulating,tang2021cutqc}. We verify this in Fig.~\ref{fig:pp}. Therefore, the number of cuts cannot be increased beyond a certain point in order to keep the classical postprocessing time in check.





\begin{figure}[htb]
    \centering
    5\includegraphics[scale=0.6]{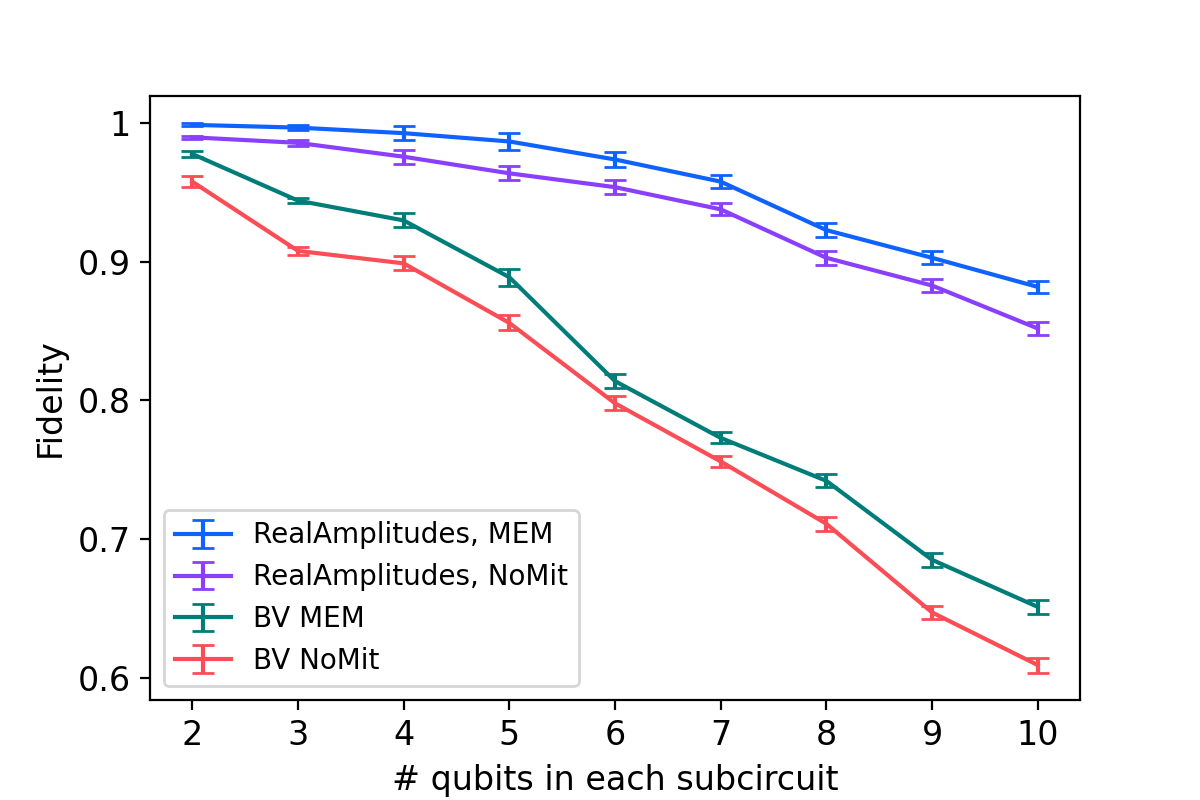}
    \caption{Fidelity obtained by the $NoTaDS$ scheduler with increasing size of the circuit where each circuit is partitioned into two subcircuits.}
    \label{fig:realampvarsize}
\end{figure}

We show next a complementary result in Fig.~\ref{fig:realampvarsize} where we increase the size of the circuit and partition each of them into two subcircuits. The two subcircuits were then scheduled by our proposed $NoTaDS$ scheduler. We notice that the fidelity decreases with an increase in the size of the circuit. The result is bootstrapped over 10 trials.

\subsection{Experimental results on 28-qubit circuit}

In our chosen set of hardware (Table~\ref{table:BACKEND}), the largest hardware has 27 qubits. Hence, here we consider one circuit that is too big to be executed on any of the hardware. However, via circuit cutting and $NoTaDS$ scheduling, we can still execute such a circuit. In Table~\ref{table: result3} we show the fidelity obtained with and without MEM for a 28-qubit RealAmplitudes circuit. We do not have any fidelity value for uncut since it is too big to be executed on our set of devices. We observe that the fidelity is poor without error mitigation, but improves significantly in the presence of MEM. This is obvious since the measurement is the most dominant noise in current quantum devices (see Table~\ref{table:BACKEND} for the probabilities of different types of noise). Therefore, the larger the circuit, the stronger the effect of measurement error, leading to poor fidelity.

\begin{table*}[htb]
\centering
\caption{Fidelity for 28-qubit circuits by scheduling over the hardware in Table~\ref{table:BACKEND} with and without circuit cutting for no error mitigation (NoMit ) and measurement error mitigation (MEM).}
\begin{tabular}{|c|c|c|c|c|c|c|c|}
\hline
\multirow{3}{*}{Benchmark circuit} & \multirow{3}{*}{\# qubits} & \multirow{3}{*}{Cut size} & \multirow{3}{*}{\# subcircuits} & \multicolumn{4}{c|}{Fidelity} \\
\cline{5-8}
& & & & \multicolumn{2}{c|}{Uncut} & \multicolumn{2}{c|}{Cut} \\
\cline{5-8}
& & & & NoMit  & MEM & NoMit  & MEM \\
\hline

RealAmplitudes &  28 & 1 & 2 & - & - & 0.31 & 0.7
\\\hline 

\end{tabular}

\label{table: result3}
\end{table*}

We have selected hardware devices up to 27-qubit devices for our experiments. Currently, IBM has hardware with 433 qubits, and our proposed method is independent of the size of the hardware.

\subsection{Change in fidelity with and without scheduling}
\label{sec:change_fid}

In all our experiments prsented so far we have fixed $\tau_j = \tau_{min}$ $\forall$ $j \in H$. Naturally, this makes the scheduling restrictive. It may be possible to execute all the subcircuits on the best device to obtain the best fidelity at the cost of execution time. Our restriction over the maximum allowable execution time $\tau$ prevented $NoTaDS$ from doing so.

\begin{figure}[htb]
    \centering
    \includegraphics[scale=0.5]{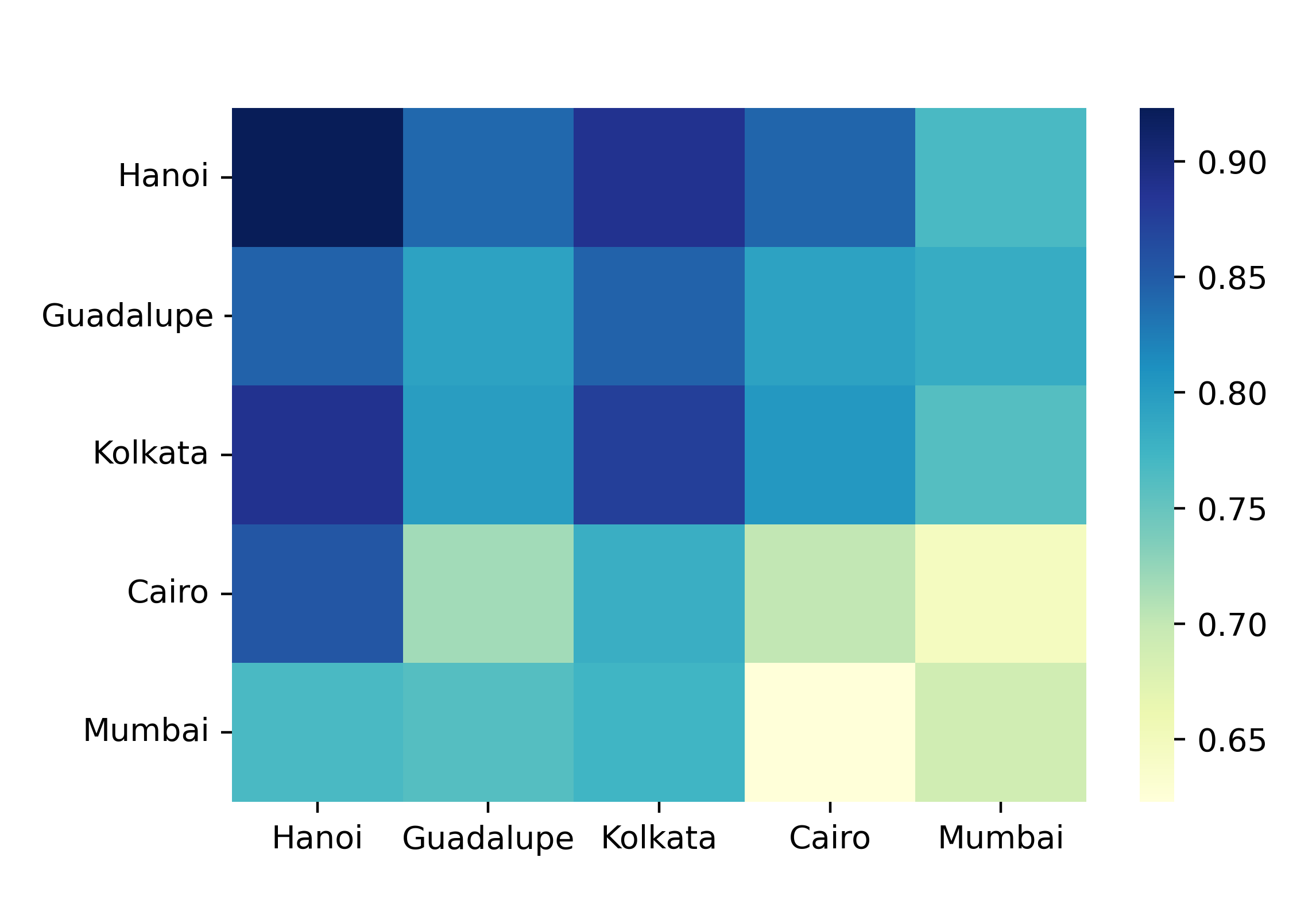}
    \caption{Fidelity of a 16-qubit RealAmplitudes circuit, when partitioned into 2 subcircuits, and executed on all possible hardware pair. Since there are only two subcircuits, when the subcircuits are scheduled to two different hardware, the execution time is $\tau_{min}$, and when scheduled to the same hardware the execution time is $\tau_{max}$.}
    \label{fig:heatmaphw}
\end{figure}

In Fig.~\ref{fig:heatmaphw} we consider a 16-qubit RealAmplitudes circuit which is partitioned into two balanced subcircuits. The number of qubits and gate count are roughly equal for both. We show the fidelity obtained when the two subcircuits are executed in all possible hardware pairs $(j,k)$, $j, k \in H$. Note that since there are only two subcircuits, we have

$$
\textit{execution time} =
    \begin{cases}
         \tau_{max} & \text{for} ~j = k \\
         \tau_{min} & \text{otherwise}.
    \end{cases}
$$

The partition being balanced, we have $\tau_{max} \simeq 2 \cdot \tau_{min}$.

The maximum fidelity obtained in Fig.~\ref{fig:heatmaphw} is when both the subcircuits are executed on \textit{ibm\_hanoi}, whereas if the two subcircuits are executed on \textit{ibm\_hanoi} and \textit{ibmq\_kolkata}, the fidelity is slightly lower but the execution time is reduced to half. The reduction in fidelity is $\sim 1\%$ only.

On the other hand, the result here indicates that if it is not possible to execute both the circuits on \textit{ibm\_hanoi} due to restrictions in execution time, it is rather more useful to schedule the subcircuits to two different hardware using $NoTaDS$ than to execute both of them together on any other hardware. This holds true even if there is some hardware whose maximum execution time can accommodate both subcircuits. For example, if we keep \textit{ibm\_hanoi} out of the story, it is better to distribute the two subcircuits to, say \textit{ibmq\_kolkata} and \textit{ibm\_cairo} than to execute both of them on the later, even if it can accommodate both. This is because \textit{ibmq\_kolkata} has a lower noise profile than \textit{ibm\_cairo}. Therefore, $NoTaDS$ finds this distributed scheduling, and improves the final fidelity of the circuit.

In Fig.~\ref{fig:heatmaphw}, the number of subcircuits is 2, so for $\tau_{max}$ both of them can be executed on the best hardware and for $\tau_{min}$ distinct devices need to be assigned. In this scenario, allowing an execution time of $\tau_{min} < \tau < \tau_{max}$ to one or more hardware cannot change the scheduling, and hence the fidelity. However, if the number of subcircuits is more than $2$, then $NoTaDS$ may be able to find even better schedules for maximum execution time $\tau_{min} < \tau < \tau_{max}$ so that the difference in fidelity obtained from the scheduling with that when all the subcircuits are executed on the best device is less than even $1\%$.

Naturally answers to the questions such as (i) what is the best schedule, (ii) is it better to schedule all the subcircuits to the same hardware -- changes with time (since noise varies with time), the list of available hardware, and the circuits. $NoTaDS$ automates this process by finding the optimum scheduling based on the hardware noise profile and the upper bound on the execution time of each hardware.

\section{A polynomial time solution for a restricted scenario}
\label{sec:matching}
In Sec~\ref{sec:ilp}, we provided an ILP solution for the scheduling problem. However, ILP is an NP-Hard problem. Therefore, in this section, we propose a graph theoretic approach for the scheduling when (i) the number of subcircuits is at most as many as the number of hardware, and (ii) the maximum execution time $\tau_j = \tau_{min}$, $\forall$ $j \in H$. Under these two restrictions, the scheduling problem essentially becomes an assignment and therefore can be solved in polynomial time using a graph theoretic approach. To differentiate this restricted scenario from the general scheduling problem, we call this \emph{circuit assignment}.

\begin{figure*}[htb]
    \centering
    \includegraphics[scale=0.28]{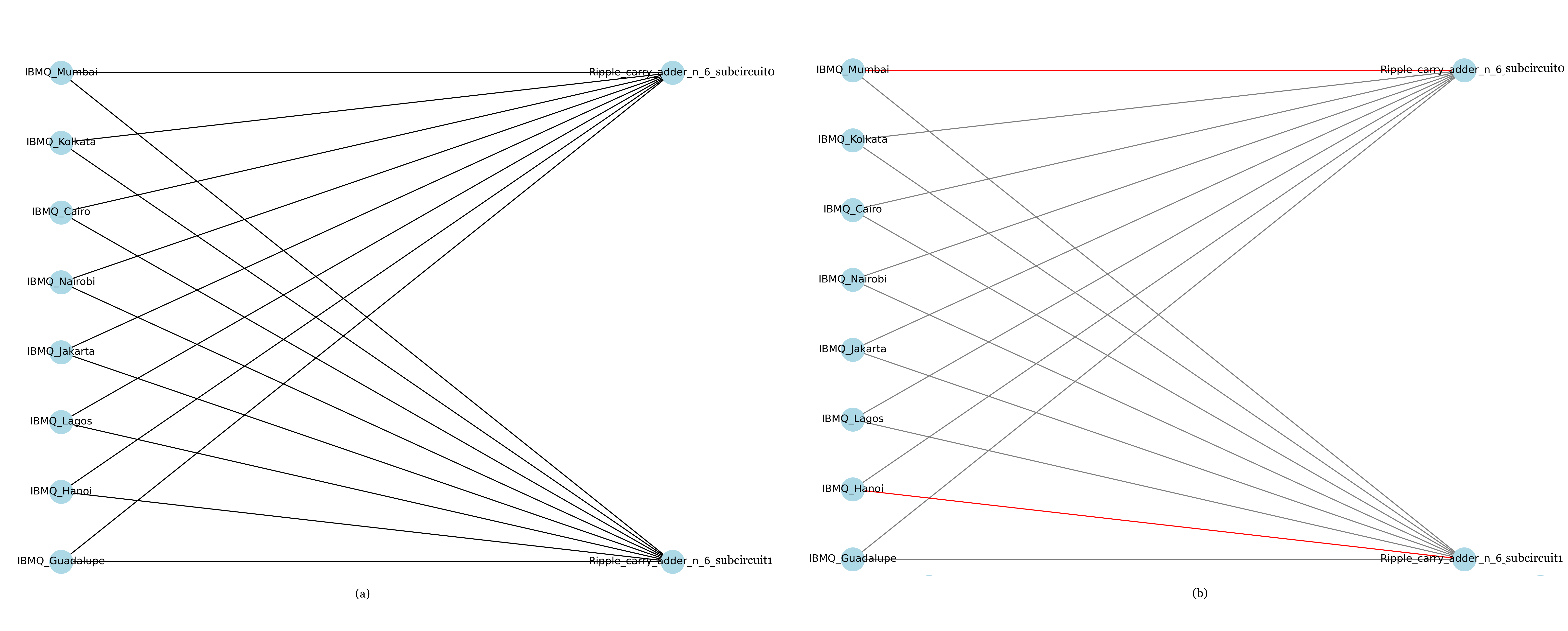}
    \caption{Mapping the 6-qubit ripple carry adder to the available hardware: (a) the bipartite graph where the  left set of vertices denotes the hardware, and the right set of vertices  the two subcircuits of a 6-qubits ripple carry adder, and 
    (b) a Minimum Weight Maximum Matching (MWMM) based solution to the assignment of two subcircuits corresponding to the 6-qubit ripple carry adder.}
    \label{fig:mwmm}
\end{figure*}
Let $C$ and $H$ be the set of subcircuits and pf hardware. We convert this \emph{circuit assignment} problem into a complete bipartite graph $G = (C, H, E)$ such that for every two vertices $c \in C$ and $h \in H$, $e = (c,h) \in E$. Such a complete bipartite graph with partitions of size $|C| = m$ and $|H| = n$ is denoted by $K_{m,n}$. As discussed in Sec~\ref{subsec:mapomatic}, we shall find the best layout and the \emph{mapomatic} score $Q_{c,h}$ for each (circuit, hardware) pair $(c,h)$. We assign a weight of $Q_{c,h}$ to each edge $e = (c,h) \in E$.
In Fig.~\ref{fig:mwmm} (a) we give an example of mapping the assignment problem to a complete bipartite graph, where the first (left) set of vertices denotes the hardware, and the second (right) set of vertices denotes the two subcircuits of a 6-qubits ripple carry adder. Each edge is associated with the \emph{mapomatic} score for the corresponding (circuit-hardware) pair. However, we have not shown the weights of the edges in the figure to keep it tidy.

In a bipartite graph, matching refers to a collection of edges selected in a manner where none of the chosen edges have a common endpoint. A maximum matching, denotes a matching that has the largest possible size, indicating the highest number of edges that can be included. It becomes a perfect matching when $|C| $ and $|H|$ are equal which is not necessarily true for our case. In our problem, we want to assign each subcircuit to hardware such that the mapomatic score is minimized. Therefore, finding the optimal noise-aware assignment of the subcircuits to the hardware is the same as finding the minimum weight maximum matching (MWMM) \cite{edmonds1965paths}.

Fig.~\ref{fig:mwmm} (b) shows the assignment of the two subcircuits to the hardware, denoted by the red edges. Note that, the ILP solution for the scheduling for the 6-qubit ripple carry adder, shown in Table~\ref{tab:adder_sched}, is also the same. Therefore, the fidelity obtained by MWMM is the same as that of the ILP scheduler.

If the number of subcircuits is more than the number of hardware, or $\tau_j > \tau_{min}$ for $j \in H$, then more than one subcircuit may be scheduled to the same hardware. This is no longer a matching, and hence cannot be solved using the MWMM method. Therefore, this polynomial time graph theoretical approach is applicable only to a restricted scenario, as discussed at the beginning of this section.

\section{Concluding Remarks} \label{sec:discussion}
In this paper, we propose a noise and time optimized distributed scheduler $NoTaDS$ that schedules the subcircuits obtained after circuit cutting to hardware such that the fidelity is maximized, and yet the execution time on each hardware is restricted by a pre-specified limit. 

We have proposed a simple method to fix this time. But this is neither the only technique by which one can estimate nor are we claiming that this is the best way to calculate it. This remains a future scope of work.

This scheduler can also be used without circuit cutting, but our method surpasses the fidelity of uncut circuits run on the least noisy devices and significantly reduces execution time. By combining inter-device parallelization with noise-aware scheduling, it optimizes circuit fidelity, which is especially useful when devices are noisy and execution time is limited. Future work may explore scheduling circuits where balanced partitioning is too costly.

\section* {Code availability}
The code to find the optimal scheduling using our proposed $NoTaDS$ scheduler is available in \href{https://github.com/debasmita2102/NoTODS}{https://github.com/debasmita2102/NoTODS}. 

\section*{Acknowledgments}
This research used resources of the National Energy Research Scientific Computing Center, a DOE Office of Science User Facility
supported by the Office of Science of the U.S. Department of Energy
under Contract No. DE-AC02-05CH11231 using NERSC award NERSC DDR-ERCAP0023266.

\section*{Conflict of interest}
The authors report no conflict of interest. Ritajit Majumdar started working on this project when he was affiliated with Indian Statistical Institute and continued the project in his current affiliation. Amit Saha also started working on this project when he was affiliated with Eviden (an Atos business) and continued the project in his current affiliation.

\bibliographystyle{unsrt}

\bibliography{main}

\end{document}